\documentclass{ws-mplb}
\usepackage{color}

\begin{document}

\markboth{H. Johannesson, D. F. Mross, and E. Eriksson}{Two-Impurity Kondo Model: Spin-Orbit Interactions and Entanglement}

%
\catchline{}{}{}{}{}
%

\title{TWO-IMPURITY KONDO MODEL: \\SPIN-ORBIT INTERACTIONS AND ENTANGLEMENT}

\author{Henrik Johannesson}
\address{Department of Physics, University of Gothenburg, SE 412 96 Gothenburg, Sweden}

\author{David F. Mross}
\address{Department of Physics, Massachusetts Institute of Technology, Cambridge, MA 02139, USA}

\author{Erik Eriksson}
\address{Department of Physics, University of Gothenburg, SE 412 96 Gothenburg, Sweden}

\maketitle 


\begin{abstract}
Motivated by proposals to employ RKKY-coupled spins as building blocks in a solid-state quantum computer, we analyze how the RKKY interaction in a 2D electron gas is influenced by spin-orbit interactions. Using a two-impurity Kondo model with added Dresselhaus and Rashba spin-orbit interactions we find that spin-rotational invariance of the RKKY interaction $-$ essential for having a well-controllable two-qubit gate $-$ is restored when tuning the Rashba coupling to have the same strength as the Dresselhaus coupling. We also discuss the critical properties of the two-impurity Kondo model in the presence of spin-orbit interactions, and extract the leading correction to the block entanglement scaling due to these interactions.

\end{abstract}


\section{Introduction}

The quest for spin-based quantum computation\cite{LossDiVincenzo} has led to a revival in the interest of the {\em two-impurity Kondo model} (TIKM)\cite{Jayprakash}, defined by the Hamiltonian
\begin{equation} \label{2impK}
H_{\text{TIKM}}= H_{\text{kin}} + J_1\boldsymbol{S}_1\cdot \boldsymbol{\sigma}_1 + J_2\boldsymbol{S}_2\cdot \boldsymbol{\sigma}_2 + K(R)\boldsymbol{S}_1\cdot\boldsymbol{S}_2.
\end{equation}
Here $\boldsymbol{S}_{1,2}$ represent two localized spins of magnitude $S=1/2$, separated by a distance $R$, and coupled to electronic spin densities $\boldsymbol{\sigma}_{1,2}$ via a Kondo interaction of amplitude $J_{1,2}$ and to each other via an RKKY interaction of amplitude $K(R)$. $H_{\text{kin}}$ is the kinetic energy of the conduction electrons.
The localized spins may be realized by two spinful quantum dots in a gated two-dimensional electron gas, with the RKKY interaction mediated by the conduction electrons in an interjacent large quantum dot, $K(R) \sim J_1 J_2 \cos(R/\epsilon)$, with $\epsilon$ a microscopic length (Fig. 1). As shown in an experiment by Craig {\em et al.}\cite{Craig}, the RKKY coupling can be controlled by adjusting the voltage on an external gate. With the two spin states on each dot representing a qubit, this suggests a means to emulate a {\em two-qubit gate}\cite{qubit}. A key issue is how robust the RKKY coupling is against competing interactions. In particular, the lack of inversion symmetry in a quantum well implies the presence of spin-orbit interactions\cite{Winkler} that may modify the simple isotropic RKKY interaction in (\ref{2impK}). As the isotropy of a spin-spin interaction ({\em alias} the coupling between qubits) is a highly desirable feature when designing a two-qubit gate\cite{Cerletti}, one is faced with the problem of engineering an {\em isotropic spin-orbit modified RKKY interaction}. Is this possible? This is the question we shall address in the first part of this article. 

The model in Eq. (\ref{2impK}) is also interesting at a more fundamental level as it is known to exhibit a quantum phase transition driven by the competition between the RKKY and Kondo interactions\cite{Jayprakash}. Adding spin-orbit interactions, this leads to a second question: How do these interactions influence the critical behavior of the model? In the second part of the paper we show how to arrive at an answer via boundary conformal field theory (BCFT)\cite{ALJ}. A fine tuned spin-orbit interaction is found to produce an irrelevant perturbation of the fixed point Hamiltonian (in the language of the renormalization group), and we shall uncover how this perturbation is encoded in corrections to the block entanglement scaling at criticality.

\begin{figure}[th]
\centerline{\psfig{file=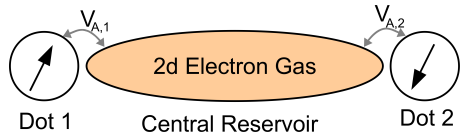,width=6cm}}
\vspace*{8pt}
\caption{(color online) Two spinful quantum dots coupled by an RKKY interaction via spin exchange with conduction electrons in a 2D central reservoir. The different $V$ are tunneling rates, related to the Kondo couplings $J_i$ in Eq. (\ref{2impK}) by $J_i \sim V^2_{A,i}/U \ (i=1,2)$, where $U$ is the Coulomb blockade energy of the reservoir. With the dots operated in the Coulomb blockade regime, charge transfer between the dots and the central reservoir are suppressed.}
\end{figure}

\section{RKKY interaction in the presence of spin-orbit interactions}

Spin-orbit interactions in a quantum well come in two brands, the {\em Dresselhaus interaction}\cite{Winkler} due to breaking of the inversion symmetry of the crystal lattice,
\begin{equation} \label{Dresselhaus}
H_{\text{D}} = \beta (k_x \sigma^x - k_y \sigma^y)
\end{equation}
and the gate-controllable {\em Rashba interaction}\cite{Winkler} coming from the two-dimensional confinement of the electrons,
\begin{equation} \label{Rashba}
H_{\text{R}} = \alpha (k_x \sigma^y - k_y \sigma^x).
\end{equation}
The amplitude ratio $\alpha/\beta$ depends on the material as well as the design and the gate bias of the particular semiconductor heterostructure which supports the quantum well, with $\alpha/\beta$ ranging from order unity in a typical GaAs/AlGaAs device to ${\cal O}(10^3)$ for HgTe/CdTe\cite{Silsbee}. Following Imamura {\em et al.}\cite{Imamura}, the RKKY interaction in the presence of the spin-orbit couplings in (\ref{Dresselhaus}) and (\ref{Rashba}) can be calculated to second order in perturbation theory as
\begin{multline} \label{pert} 
H_\text{RKKY}=
-\frac{J_1 J_2}{\pi}\text{Im}\int_{-\infty}^{\omega_F}
d\omega \ \text{Tr}\, \big[ (\boldsymbol{S}_1\cdot\boldsymbol{\sigma})G({R/2},\omega+\i 0_+)\\ \times (\boldsymbol{S}_2\cdot\boldsymbol{\sigma})G(-{R/2},\omega+\i 0_+) \big], 
\end{multline}
with $\omega_F$ the Fermi energy.
Here $G({R/2},\omega)$ is the Green's function of a conduction electron with single-particle Hamiltonian $H_{\text{el}}=H_{\text{kin}}+H_{\text{D}}+H_{\text{R}}$, and with the trace in (\ref{pert}) taken over its two spin states. By choosing a coordinate system where the vector $R \hat{R}$ that joins the two quantum dots is parallel with the $\hat{x}$-axis, a lengthy calculation yields that\cite{MJ}
\begin{equation}H_\text{RKKY}=H_{0}+H_{\alpha}+H_{\beta}+H_{\alpha \beta}, \end{equation}
where
\begin{equation}
 		\begin{array}{clcl}
&H_{0}&=&F_{0}\boldsymbol{S}_1\cdot\boldsymbol{S}_2\\[1mm]
&H_{\alpha}&=&\alpha F_{1}\left(\boldsymbol{S}_1\times\boldsymbol{S}_2\right)^y+\alpha^2F_2S_1^yS_2^y\\[1mm]
&H_{\beta}&=&\beta F_{1}\left(\boldsymbol{S}_1\times\boldsymbol{S}_2\right)^x+\beta^2F_2S_1^xS_2^x\\[1mm]
&H_{\alpha \beta}&=&\alpha\beta F_2\left(S_1^xS_2^y+S_1^yS_2^x\right).\label{eq:terms}\end{array}
\end{equation}
The functions $F_i=F_i(\alpha,\beta,R), i=0,1,2$ are given by rather complicated integrals which in the the general case must be calculated numerically.

When turning off the spin-orbit interactions in (\ref{Dresselhaus}) and (\ref{Rashba}), i.e. with $\alpha \!= \!\beta\! = \!0$ in (\ref{eq:terms}), one obtains the standard spin-rotational invariant form of the RKKY interaction in (\ref{2impK}), with $K(R) \!= \! F_0$. To find out whether there are any finite values of $\alpha$ and $\beta$ for which spin-rotational invariance may be recovered, it is useful to rotate the coordinate system by an angle $\arctan(\alpha/\beta)-\pi/2$ around the $\hat{z}$-axis. The spin-orbit modified RKKY interaction then takes the form
\begin{eqnarray} 
H_{\text{RKKY}}&=&K_\text{H} \boldsymbol{S}_1\cdot\boldsymbol{S}_2 +K_\text{Ising}S_1^y S_2^y +K_\text{DM}\left(\boldsymbol{S}_1\times\boldsymbol{S}_2\right)^y,   \label{RKKYinter}
\end{eqnarray}
with $K_{\text{H}}, K_\text{Ising}$, and $K_{\text{DM}}$ parameterized by $\alpha, \beta, J_1, J_2,$ and $R$. The Ising and Dzyaloshinski-Moriya (DM) terms in (\ref{RKKYinter}) can be traced back to a term in the Green's function $G(R/2, \omega)$ in (\ref{pert}) which contains a factor $\mathbb{A} \hat{R}$, with 
$\mathbb{A}$ a matrix with elements $A_{11}\!=\!-A_{22}\!=\!\beta, A_{12}\!=\!-A_{21}\!=\!-\alpha$. In the rotated coordinate system one has that $\mathbb{A}\hat{R}=\alpha\left(0,(\alpha^2-\beta^2)\cos\arctan(\alpha/\beta)\right)$ and it follows from (\ref{RKKYinter}) that the RKKY interaction becomes manifestly spin-rotational invariant when $\alpha=\beta$. Our result boosts the proposal in Ref. 3 that an RKKY-coupled double-quantum dot device can be used as a building block of a two-qubit gate, {\em also in the realistic case with spin-orbit interactions included}. Anisotropic terms in a pulsed spin exchange used in a two-qubit gate are well-known to be a nuisance, as they tend to mix different spin states, implying a slow-down of the switching time\cite{Cerletti}. While various schemes to get around or reduce this problem have been proposed\cite{Bonesteel,BurkardLoss}, the ideal two-qubit gate is patterned upon an isotropic exchange. As revealed by our results, this situation should in principle be possible to achieve experimentally in a GaAs or InAs quantum well where the Rashba coupling $\alpha$ is of comparable strength to that of $\beta$ and can be fine-tuned by an external gate voltage.  

Equal strengths of the Rashba and Dresselhaus interactions have been found to restore also a type of ``hidden'' SU(2) symmetry in a 2D electron gas, predicted to produce a {\em persistent spin helix} 
$-$ a helical spin density wave of infinite lifetime\cite{BAB}. It would be interesting to explore whether there is a connection between this finding and that of ours for the RKKY interaction. For related work on implications of equal strengths of Dresselhaus and Rashba couplings, see Refs. 11 and 12.

\section{Quantum criticality with spin-orbit interactions}
The results in the previous section are valid only when the direct Kondo interactions in (\ref{2impK}) are dominated by the RKKY term, i.e. when $K(R) \gg T_K$, where $T_K$ is the energy scale (``Kondo temperature'') below which Kondo screening sets in\cite{Coleman}. By tuning the gate voltages so that $T_K$ becomes larger than $K(R)$ one passes into a {\em Kondo phase} where the two localized spins are completely screened. When the electrons that mediate the RKKY interaction are separated from those that participate in the Kondo screening, the system undergoes a quantum phase transition with a  non-Fermi liquid quantum critical point\cite{Chung}. The condition of electron separation can be realized in the laboratory by using a setup as in Fig. 2. By proper gating, the electrons in the central dot are made to mediate an RKKY interaction only, with the external leads providing for the possible Kondo screening channels.\cite{discussion} As shown by Affleck {\em et al.}\cite{ALJ}, an efficient way to characterize the quantum critical behavior of the system is to use a BCFT approach. Its extension to the case with Dresselhaus and Rashba interactions included in the external leads meets with some technical difficulties, however, connected to the fact that there is now an intertwined sequence of orbital angular modes coupling to the localized spins. To handle this situation one needs some powerful scheme, with details still to be worked out\cite{Zitko}. However, when only {\em one} type of spin-orbit interaction in the external leads is present (i.e. Dresselhaus {\em or} Rashba, but as before allowing for both types of interactions to be simultaneously present in the central dot), the BCFT approach still delivers very effectively. For this case the individual spin-orbit interactions connect only two angular momentum modes to the Kondo coupled s-wave component of the electron fields, thus simplifying matters dramatically\cite{Malecki}.

\begin{figure}[th]
\centerline{\psfig{file=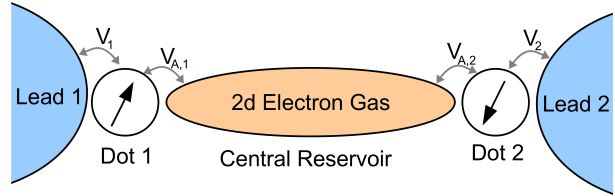,width=8cm}}
\vspace*{8pt}
\caption{(color online) The double-quantum dot system with central electron reservoir and  attached leads. The different $V$ are tunneling rates. The dots are operated in the Coulomb blockade regime, where charge transfer between the reservoir and dots as well as between the leads and the dots is strongly suppressed. By proper gating, electrons in the reservoir [leads] will mediate [participate] in the RKKY interaction [Kondo screening] only.}
\end{figure}

In order to apply the BCFT machinery it is convenient to first rotate the localized spin on {\em one} of the dots, $\boldsymbol{S}_2 \rightarrow \boldsymbol{S}_2^{\prime}$, followed by the same rotation of the electron spins in the lead connected to this dot, $\boldsymbol{\sigma}_2 \rightarrow \boldsymbol{\sigma}_2^{\prime}$. By a judicious choice of twist angle $\theta$ (for details, see Ref. 9), the full {\em spin-orbit modified TIKM} can be cast on the form
\begin{equation} \label{SOTIKM}
H= H_{\text{kin}} + J_1\boldsymbol{S}_1\cdot \boldsymbol{\sigma}_1 + J_2\boldsymbol{S}^{\prime}_2\cdot \boldsymbol{\sigma}^{\prime}_2 + K^{\perp}\boldsymbol{S}_1 \cdot \boldsymbol{S}^{\prime}_2 + (K^y-K^{\perp})S_1^yS_2^{\prime y},
\end{equation}
where $K^{\perp}$ and $K^y$ are parameterized by $K_H, K_{\text{Ising}}, K_{\text{DM}}$ and $\theta$, and where, for simplicity, we have chosen $J_1=J_2 \equiv J$. When $K^y=K^{\perp}$  we recover the ordinary TIKM in (\ref{2impK}), and for this case the critical behavior is therefore the same for all twist angles $\theta$. The case where there is no twist, but an Ising anisotropy, $K^y\neq K^{\perp}$, is a bit different. Now the SU(2) symmetry of the theory is broken down to U(1). We know from Ref. 6 that anisotropies in the Kondo interaction do not change the leading scaling behavior at criticality. Now, whether the symmetry breaking is due to an anisotropy in the Kondo exchange, or, as in (\ref{SOTIKM}), in the RKKY interaction, is immaterial since the BCFT operator content which governs the critical behavior depends only on the overall left-over symmetry. One may thus be tempted to conclude that the renormalization-group fixed point that governs the critical theory is stable also against an Ising anisotropy in the RKKY interaction. However, this line of argument would be too fast: Changing the parameters $K^{\perp}$ and $K^y$ by the {\em same} amount is in fact a relevant perturbation, taking the scaling Hamiltonian away from the fixed point along an SU(2)-invariant direction. Given the linearized RG flow around the fixed point, the irrelevant direction is perpendicular to the SU(2) invariant line, and thus, only by fine-tuning the parameters by letting $K^{\perp} \rightarrow K^{\perp} + \delta$ and $K^{y} \rightarrow K^{y} - \delta$ will we stay at the fixed point.
Since the fixed points for all values of the twist $\theta$ arise from the same Hamiltonian and should thus be identified, it follows that the general case with both a twist {\em and} an Ising anisotropy is the same as that with no twist. In other words, the TIKM fixed point is stable under perturbations of spin-orbit interactions provided that one tunes the parameters judiciously,  $K^{\perp} \rightarrow K^{\perp} + \delta$, $K^{y} \rightarrow K^{y} - \delta$.  Else one flows towards one of the stable fixed points representing the RKKY phase and the Kondo screened phase, respectively (see Fig. 3).

\begin{figure}[th]
\centerline{\psfig{file=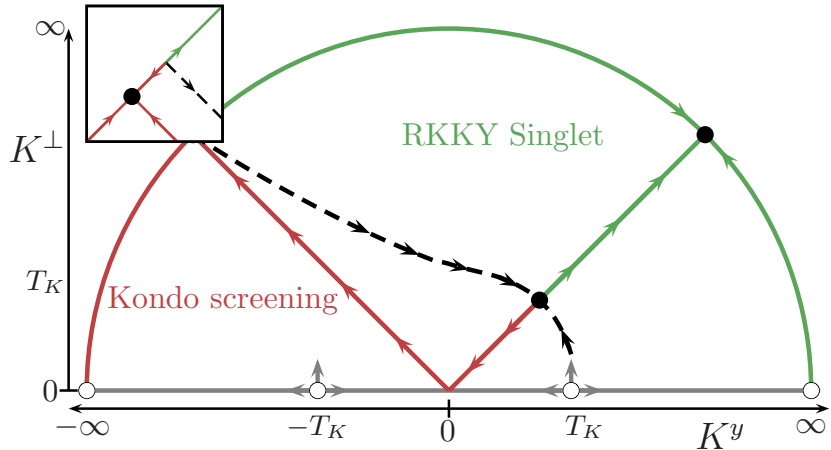,width=11cm}}
\vspace*{8pt}
\caption{(color online) RG flow of the spin-orbit modified TIKM. The solid dots and the solid line are known  results for the ordinary TIKM with no spin-orbit interactions (Ref. 6). The gray line, with $K^{\perp}=0$, marks the RG flow of a different  model of quantum dots coupled via an Ising interaction where different behavior is expected (Ref. 18). The dashed flow line separates the RKKY singlet from the Kondo screened regime.  As an artifact of the scale at $|K|\rightarrow \infty$ this line appears curved to coincide with the screened fixed point ($K^\perp=-K^y$). Note that at both the RKKY and the Kondo screened fixed point, the direction along the semicircle is irrelevant. We thus expect there to remain a finite separation (its scale being set by $T_K$)  between the relevant flow towards the Kondo screened fixed point and the dashed flow towards the critical point, as  shown in the enlarged inset. The curvature of the dashed line is not meant to suggest any deeper knowledge about its properties; however, close to the isotropic (unstable) fixed point it follows the direction of irrelevant longitudinal anisotropies.}
\end{figure}

\section{Entanglement at criticality}
When considering proposals for quantum information processing using RKKY-coupled spins\cite{Craig}, it becomes interesting to quantify the entanglement between the two spins as measured by the {\em concurrence}\cite{review}. This was done by Cho and McKenzie\cite{ChoMcKenzie} who also showed that the concurrence vanishes identically at the critical point and thus serves as a marker for the quantum phase transition. This result remains valid in the presence of spin-orbit interactions\cite{MJ}. 

Another entanglement measure with interesting scaling behavior at a quantum phase transition is the {\em block entanglement}\cite{review}. For a system in a pure state and partitioned into two parts $A$ and $B$, the block entanglement is encoded by the {\em von Neumann entropy} $S_A = - \mbox{Tr}\,\rho_A \log\rho_A$ of the reduced density matrix $\rho_A$ (with $S_A = S_B$). The contribution to 
$S_A$ coming from the impurity spins in the TIKM Hamiltonian in (\ref{2impK}) is given by the {\em boundary entropy} $s_{\text{b}} = \log\sqrt{2}$ at the nontrivial fixed point \cite{ALJ}. Since the critical properties of the spin-orbit modified TIKM in (\ref{SOTIKM}) are controlled by the same fixed point as for the TIKM without spin-orbit interactions, it follows that the boundary entropy stays the same. To uncover the presence of the spin-orbit interactions in (\ref{Dresselhaus}) and (\ref{Rashba}) one must study the corrections to 
$s_{\text{b}}$ implied by the enlarged content of RG-irrelevant operators due to the breaking of SU(2) 
$\rightarrow$ U(1) when $K^y \neq K^{\perp}$. A symmetry analysis suggests that the boundary operator corresponding to the irrelevant RG flow in Fig. 3 can be identified as a component of the energy-momentum tensor, with scaling dimension $x_{\text{b}}\!=\!2$. By employing a formalism developed by Cardy and Calabrese for calculating corrections to the critical von Neumann entropy\cite{CC}, recently adapted to the case of {\em boundary perturbations}\cite{EJ}, we then infer that
\begin{equation} \label{correction}
s_{\text{b}} = \log\sqrt{2} + \frac{\xi_{\text{K}}}{r}(a \log(\frac{r}{\xi_{\text{K}}}) + b) + ...  
\end{equation}
where $a$ and $b$ are dimensionless constants, $\xi_{\text{K}}$ is the {\em Kondo screening length},\cite{Coleman} and where $2r >> 2 \xi_{\text{K}} >> \ell$ is the length of the block, 
with $\ell$ being the distance between the dots (see Fig. 4). Higher-order corrections are denoted by $``...`$. Here the $\log(r)/r$ term is produced by the {\em leading} irrelevant boundary operator of dimension $x_{\text{b}}=3/2$, with the second term in the parenthesis contributed precisely by an $x_{\text{b}}\!=\!2$ operator. It is important to point out that a $1/r$-term appears also for the TIKM in (\ref{2impK}) with no spin-orbit interactions, due to the fact that the {\em full} energy-momentum tensor is part of {\em any} scaling Hamiltonian. Thus, to leading order, the presence of the spin-orbit interactions is revealed only by a change of the amplitudes $a$ and $b$ in Eq. (\ref{correction}). 
\begin{figure}[th]
\centerline{\psfig{file=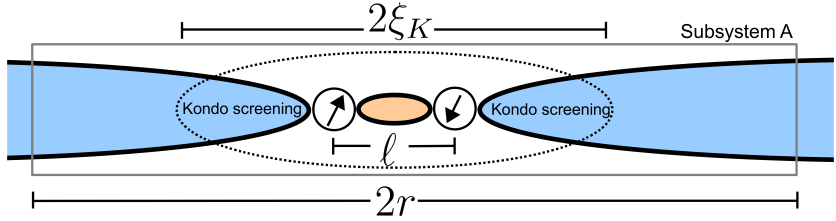,width=11cm}}
\vspace*{8pt}
\caption{(color online) Schematic of the double-quantum dot system with a block A of length $2r \gg 2\xi_{\text{K}} \gg \ell$ (see text).} 
\end{figure}

\section{Summary}

We have investigated the influence from Dresselhaus and Rashba spin-orbit interactions on the RKKY coupling between two impurity spins in a 2D electron gas. By proper gating, the Dresselhaus and Rashba coupling strengths can be made equal, for which the RKKY interaction is found to become manifest spin-rotational invariant. As this property is a {\em sine qua non} for employing RKKY-coupled spins for two-qubit gating, our result adds to the viability of the scheme also in the presence of spin-orbit interactions.

We have also explored how the Dresselhaus and Rashba interactions influence the quantum critical behavior of the two-impurity Kondo model. By fine-tuning their coupling strengths the system can be made to stay critical, being governed by the same fixed point as for the model without spin-orbit interactions. As a result, the impurity contribution $\log\sqrt{2}$ to the block entanglement remains the same, with the presence of the spin-orbit interactions showing up only in the amplitudes of the subleading scaling corrections. It would be interesting to widen the search for spin-orbit effects on quantum impurity critical behavior, for example by including the occurrence of charge fluctuations {\em (two-impurity Anderson model)}\cite{MJ_0}, adding more interaction channels {\em (Kondo quartet model)}\cite{Georges}, or by adding a third impurity spin {\em (Kondo trimer model)}\cite{Fabrizio}.

\section*{Acknowledgments}

We wish to thank P. Calabrese, C.-H. Chung, H. R. Krishna-murthy, and Y. Meir for discussions and helpful correspondence. This research was supported by the Swedish Research Council under Grant No. VR-2008-4358 (HJ).

\end{document}